# PREDICTION OF TISSUE DECOMPRESSION IN ORBITAL SURGERY

Vincent Luboz[1], Annaig Pedrono[2], Dominique Ambard[2], Frank Boutault[3], Yohan Payan[1], Pascal Swider[2]

1. TIMC/IMAG Laboratory, UMR CNRS 5525, 38706 La Tronche, France
2. Biomechanics Laboratory, EA 3697, and 3. Maxillofacial Department, Purpan University Hospital, 31059 Toulouse, France

Corresponding author:
Luboz Vincent
Laboratoire TIMC/IMAG,
UMR CNRS 5525,
Institut d'Ingénierie de l'Information de Santé
Faculté de Médecine Domaine de la Merci
38706 La Tronche
France

Tel: +33 4 76 52 00 10
Fax: +33 4 76 52 00 55
e-mail: Vincent.Luboz@imag.fr

**Abstract**

*Objective*:

A method to predict the relationships between decompressed volume of orbital soft tissues, backward displacement of globe after osteotomy, and force exerted by the surgeon, was proposed to improve surgery planning in exophthalmia reduction.

*Design*:

A geometric model and a poroelastic finite element model were developed, based on Computed Tomography scan data.

*Background*:

The exophthalmia is characterised by a protrusion of the eyeball. Surgery consists in an osteotomy of the orbit walls to decompress the orbital content. A few clinical observations ruling on an almost linear relationship between globe backward displacement and tissue decompressed volume are described in the literature.

*Methods*:

Fast prediction of decompressed volume is derived from the geometric model: a sphere in interaction with a cone. Besides, a poroelastic Finite Element model involving morphology, material properties of orbital components and surgical gesture was implemented.

*Results*:

The geometric model provided a better decompression volume estimation than the Finite Element model. Besides, the Finite Element model permitted to quantify the backward displacement, the surgical gesture and the stiffness of the orbital content.

<u>*Conclusions*</u>:

The preliminary results obtained for one patient, in accordance with the clinical literature, were relatively satisfying. An efficient aid for location and size of osteotomies was derived and seemed to be able to help in the surgery planning.

**Relevance:**

To our knowledge, this paper concerns the first biomechanical study of exophthalmia reduction. The approach permitted to improve the treatment of orbitopathy and can be used in a clinical setting.

# 1. Introduction

Acquired exophthalmia is frequently observed in cases of thyroid diseases and especially in cases of Grave's disease which is associated with hyperthyroidy (Saraux *et al.*, 1987). In these cases, it can be observed an excessive protrusion of the ocular globe outside the orbit, due to an increase of volume of the orbital content (fat tissues and ocular muscles), Figure 1a. Besides its aesthetical and psychological consequences, the exophthalmia is often followed by an abnormal exposition of the cornea and by the extension and/or the compression of the optic nerve and orbital blood vessels. Those factors can lead to a decrease of visual acuity and sometimes to total blindness. At worst, exophthalmia needs active therapy and often surgery to avoid serious complications. First the endocrinal dysfunction has to be stabilised with medicines or radiotherapy in the worst cases. Then surgical reduction of the exophthalmia is usually needed to decompress the orbital content.

Two surgical techniques can be used for this decompression. The first one aims at extracting a partial volume of fat tissues via an external eyelid incision (Olivari, 2001). The advantages of this method are its simplicity and the relative security during surgery since the optic nerve is far from the extracted fat tissues. The drawbacks are first a small backward displacement of the globe due to the limited volume of fat tissues in the eyelid region, the difficulty to know the exact amount of fat to be removed and an aesthetical risk for the patient since upper lid depression can be later observed. The second surgical technique aims at increasing the volume of the orbital cavity thanks to an osteotomy of one or several of the orbital walls (Wilson and Manke, 1991, Stanley *et al.*, 1989). These osteotomies are completed by some cuts in the membrane containing the orbital soft tissues, in order to permit a fat tissue issue. An external pressure is additionally exerted by the surgeon onto the eyeball, during a few seconds. Consequently, some retro-ocular tissues can be evacuated through the hole, towards the sinuses. The advantages of this method are the conservation of the integrity of intra conic fat, a backward displacement of the eyeball that can be larger than with the other technique and the theoretical possibility, for the surgeon, to have a better prediction of the post-operative result. The drawback is the risk to cause perturbations in visual function (e.g. transient diplopia). The surgical procedure has therefore to be very precise to avoid critical structures such as the optic nerve. Moreover this intervention is technically difficult since the eyelid incision is narrow and the operating field restrained; the surgery must be mini-invasive and specific tools to improve surgical planning would be really useful.

Up to now, very few results focusing on exophthalmia reduction appear in the literature. From a clinical point of view, the prediction of the surgically decompressed volume lies on observations and statistical analyses (Adenis and Robert, 1994): for a 1 $cm^3$ soft tissues decompression, a backward displacement from 1 mm to 1.5 mm is expected . From a biomechanical point of view, there are some works concerning the modelling of the ocular globe (Sagar *et al.*, 1994, Uchio *et al.*, 1999) and the intra orbital soft tissues (Miller and Demer, 1999, Buchberger and Mayr, 2000), but none of them deals with exophthalmia reduction.

To improve surgery planning in exophthalmia reduction a method to predict the relationship between the decompressed volume of orbital soft tissues and backward displacement of the orbital globe after osteotomy has been developed. In this paper, two complementary predictive models are proposed. First, a geometrical 2D model provides a rough evaluation of the decompressed tissue volume and second, a 3D poroelastic finite element model permits to simulate the surgical gesture. The clinical application concerns a patient suffering from a bilateral exophthalmia, whose only the left eye has been treated (Figure 1b).

## 2. Methods

*2.1 Three-dimensional reconstruction of the orbit*

Horizontal Computed Tomography (CT) slices were collected for the whole patient skull (helical scan with slices reconstructed each millimeter). CT images were first segmented to extract the bone limits, the muscles and the optic nerve contours. The segmentation procedure was achieved manually through B-Splines definitions on each slice, as shown in Figure 1b. The area of each segmented anatomical structure were deducted and then extrapolated over the whole slices to get the global volume. In-house software available for the surgeon has been developed in Visual C++ coupled with VTK graphical library (http://public.kitware.com/VTK/) and permitted a first 3D description of the orbital content.

*2.2 Geometrical model*

The eyeball was modelled by a sphere (radius $r_0$) lying on a homogenous cone (the orbital cavity; depth $h$, radius of opening $r$), see Figure 1c. During the decompression, the soft tissues were supposed to be constrained by the eyelid so that their extremity in this region is constant relative to the eyeball position ($x_0$). This assumption permitted to derive equation (1) expressing the relationship between the tissue-decompressed volume $V$ and the backward displacement $x$ of the eyeball.

$$V = \frac{\pi}{3}(r/h)^2(x^3 - 3x^2h + 3h^2x) \tag{1}$$

The sensitivity of parameters *r* and *h* were quantified using a perturbation technique. The first order Taylor expansion of $V_{geo}$ led to the relative variation $\Delta V/V$ of the decompressed volume expressed by equation (2).

$$\frac{\Delta V}{V} = S_r \frac{\delta r}{r} + S_h \frac{\delta h}{h} + S_{rh} \frac{\delta r}{r} \frac{\delta h}{h} + O(r^2, h^2) \qquad (2)$$

with $S_r = 2 \quad S_h = \dfrac{(3h - 2x)x}{3h^2 + (x - 3h)x}$ and $S_{rh} = 2 S_h$

The depth *h* varied from 30 to 50 mm and the backward displacement *x* was between 0 and 10 mm. Usually, the average backward displacement obtained in a clinical setting is around 5 mm. It was demonstrated by equation (2), that the sensitivity to *r* was constant ($S_r = 2$) and predominant by comparison with $S_h$ (bounded to 0.37 for *h* = 30 mm and *x* = 10) and $S_{rh}$.

*2.3. Poroelastic Finite Element model of the orbit*

The eye ball was considered as rigid in comparison with the orbital soft tissue (fat, muscles and optical nerve). From a biomechanical point of view, its strain energy was therefore neglected in the model. In this preliminary study, the orbital content was considered as an homogenized material modelling the fat tissues, the muscles and the optic nerve. The fat tissues are predominant (approximately 3/4 of the volume) in this material and are mainly present in the flow through the osteotomy during the surgical intervention. Clinical observations (Saraux *et al.*, 1987) describe the intra-orbital pathological fat tissues as the combination of an elastic phase composed of fat fibres (mainly collagen) and a fluid phase composed of fatty nodules saturated by physiological fluid.

The biomechanical modelling of the intra orbital tissue was therefore achieved using a homogeneous poroelastic formulation of the finite element (FE) method, with fluid transport in saturated elastic porous media (FE software MARC©, MSC Software Inc., Santa Ana, CA, USA). Two parameters of the poroelastic model, namely the porosity and the permeability, allow to take into account the retention of fluid, generally observed in clinical settings. The variational formulation of the coupled problem led to the finite element discretisation using nodal displacement and pore pressure as nodal variables (Biot, 1941).

The FE orbital cavity was meshed using three-dimensional 20-node hexahedrons (quadratic elements) involving three translation degrees of freedom for each node and a pore pressure for each corner node. The global meshing involved 6948 nodes, 1375 elements and 22716 degrees of freedom. The Young modulus of the homogenized tissue was set to 20 kPa according to the soft tissue properties from the literature (Fung, 1993, Power *et al.*, 2001). The Poisson ratio, set to 0.1, was inherent to the poroelastic modelling of soft tissue (Mow *et al.*, 1980, Aroubi and Shizari-Adl, 1996).

The relaxation time of the fluid saturated porous media is highly related to the permeability. In intervertebral disk, complete relaxation occurred at night in almost 8 hours and, in exophthalmia reduction, the relaxation time is around 10 seconds. A proportional rule using the relative permeability of disk as a reference ($k/\mu$ = 0.1 mm$^4$/N.s, Simon *et al.*, 1985) permitted to roughly estimate the permeability of the orbital content around 300 mm$^4$/N.s. The porosity of the homogenized orbital content was set to 0.4, also referring to the disk (Simon *et al.*, 1985).

The mesh and the boundary conditions are shown in Figure 2. Specific boundary conditions were defined to reproduce the surgical technique. Before the bony walls osteotomy, the tissues were confined in the orbit: the boundary conditions were nil displacement and total sealing effects at interface nodes. To model the bony wall osteotomy, nodes were fixed in terms of displacement to simulate the rigidity of the membrane surrounding the tissues, and they were released in term of sealing boundary conditions, to simulate cuts in this membrane due to the surgical gesture. The wall opening was located along the axis of the orbit in terms of percentage of the depth $h$. The osteotomy corresponding to the clinical data (derived from the post operative CT scan) was located on the front of the orbit, at $h_\%=62$, and had an area of 2.9cm².

To study the sensitivity of the model according to the size and the location of the wall opening, three other osteotomies were designed. Two locations were considered $h_\% = 50$ (back of the orbit) and 62 (front of the orbit) as shown in Figure 2. In accordance with clinical feasibility, two sizes were considered in the ethmoid sinuses: medium osteotomy: 1.4 cm² and large osteotomy: 2.9 cm², see Figure 2.

Riemann *et al.* (1999) have shown that the intra-orbital tissues are characterized by an overpressure of 10kPa due to the pathology. Therefore, an initial pore pressure was set to 10 kPa in the FE model to take into account the initial non-nil strain energy in the biological structures. This initial pressure was relative to the pressure applied to the orbit and meant that there was an overpressure of 10 kPa in the tissues. A relaxation time of 2 seconds was then needed to get pressure equilibrium. During this time, nodes located at the soft tissue/globe interface were free to move.

Since the eyeball is considered as a solid body, the force imposed by the surgeon was modelled as an imposed axial load (according to the long axis of the orbit) applied on an external node connected by rigid elements to the nodes of the tissue/globe interface (Figure 2). This permitted to simply control the load on interface nodes via the resulting load applied on the external node. This imposed load was linearly applied in 2 seconds in the range 0 N – 12 N as shown in Figure 3. No sliding and friction effects were considered in tangential directions and consequently, no contact analysis was needed. The backward displacement of the globe was derived from the FE computation tracking the displacement of the external node where the imposed load was applied.

The imposed load was maintained during 3s, and then released. During the release phase, the osteotomy wall was maintained impermeable to avoid fluid flow back into the orbit. This phenomenon is clinically relevant as the fat tissues stay in the sinus after having been decompressed.

The tissue-decompressed volume was derived from the difference of fluid saturated mesh volumes before and after loadings. It was corresponding to the volume of outward fluid flow. The volume of solid phase was not taken into account in the decompressed volume computation since nil displacement were imposed as boundary conditions on nodes of the osteotomy wall.

## 3 - Results

The clinical results of the exophthalmia reduction, measured from the patient CT scans as described in paragraph 2.1, were respectively 3.6 cm$^3$ for the decompressed volume and 2.9 cm$^2$ for the sinuses wall osteotomy (±14% due to errors measured on 15 segmentation cycles repeated by a single user). The central zone of the osteotomy was located at $h_\%$ = 62. The backward displacement of the eyeball was about 4 mm. The magnitude of load of 12 N applied by the surgeon on the eye ball (and used as imposed load in the FE model) was measured using a custom designed sensor developed in the laboratory.

The 3D reconstruction of the orbital components permitted to quantify the radius of the orbit opening, $r$=18 mm and the depth, $h$=42 mm. These values, introduced in equation (1), gave the tissue decompressed volume $V_{geo}$, expressed by equation (3). The tissue-decompressed volume $V_{fem}$ derived from successive FE computations was linear and was described by equation (4) for the large size osteotomy (2.9 cm$^2$) located at 26 mm ($h_\%$=62) and corresponding to the wall opening reconstructed from the post operative CT scan.

$$V_{geo} = a_1\ x - a_2\ x^2 + a_3\ x^3 \qquad (3)$$

$$V_{fem} = b_1\ x \qquad (4)$$

where V is in cm$^3$ and $x$ is in cm, $a_1$ = 1.018 cm$^2$, $a_2$ = 2.4.10$^{-2}$ cm, $a_3$ = 2.10$^{-4}$ unless dimension and $b_1$ = 0.575 cm$^2$.

Both volumes $V_{geo}$ and $V_{fem}$ were close to a linear function of the displacement $x$ since the significant coefficients were respectively 1.018 for the geometrical model and 0.575 for the poroelastic FE model (the coefficient for $x^2$ and $x^3$ are far lower). This result is highlighted in Figure 4. The geometric model provided satisfying predictive results. Indeed, it gave a volume of 3.69 cm$^3$, which is very close to the clinical result (3 % greater). The FE model with a frontal (at $h_\%$=62) large size osteotomy (derived from the clinical data) gave an estimation of the decompressed volume of 2.3 cm$^3$, 36 % lower than the clinical result.

Table 1 summarizes those results, and give a comparison of the backward displacement of the two models. Since the backward displacement is an input of the geometrical model, there is no backward displacement difference between this model and the clinical measurements. Concerning the FE model, the backward displacement discrepancy with the clinical result was very small (2 %). Differences between the two models were much more important in terms of decompressed volume: while the geometrical model was quite close to clinical data (a 3% discrepancy), the FE model under-estimated the volume (36% lower than clinical measurements).

Table 1 also plots the stiffness values for the clinical case and for the FE case. In vivo per-operative measurements of the surgical gesture are under development. The preliminary measurements show that for an average load of 12 N, a displacement of 5 mm is obtained, which corresponds to an orbital content stiffness of 2.4 N.mm$^{-1}$. This value was measured while the load is maintained constant, so that the displacement is maximal. The predicted stiffness derived from the FE model was, at this point, 2.85 N.mm$^{-1}$ (load: 12 N, displacement: 4.34 mm). Stiffness discrepancy was therefore around 19 % in that case.

Table 2 presents the results for the four osteotomies (2 sizes and 2 locations) simulated with the FE model for an imposed load of 12 N. The backward displacements, the decompressed volumes and the stiffness presented in this table point out the influence of the size and the location of the osteotomy. Indeed, the backward displacement measured for the four cases showed significant differences: from -17 % to -33 % compared to the large frontal osteotomy simulated with the Finite Element model. Concerning the decompressed volume, the variations of the results for the four osteotomies range between -3 % and -28 %. The force - displacement relationships, i.e. the stiffness, for the four cases, is time dependant and is plotted in Figure 5. Its maximum value (at t = 4 s) was used in Table 2, and range between -24 % and +48 % compared to the large frontal osteotomy.

## 4. Discussion

The geometrical model is a powerful tool for a first estimation of the volume of decompressed tissue in the pre-operative diagnosis, and it seems in good agreement with empirical clinical observations (Adenis and Robert, 1994, Jin *et al.*, 2000, ratio ≈ 1 between eye ball backward displacement and tissue decompressed volume). As demonstrated by the sensitivity analysis developed in paragraph 2, the radius of the orbit opening seemed to be a parameter requiring attention during the CT scans segmentation.

But the geometric approach did not permit to predict the impact of the surgical technique: size and location of the wall osteotomy, and clinical feasibility of the force exerted by the surgeon. These two points were addressed by the poroelastic FE model.

The FE simulation of the surgery (with the large size osteotomy at $h_\%$=62) provided a quite precise estimation of the globe backward displacement but with a significant error in the simulated decompressed volume. From our point of view, those results must be considered with caution. Indeed, the FE parameters ($E$, $\nu$ and $k$) were chosen in order to fit data and to be still in accordance with rheological data published in the literature. This means that those parameters must be more extensively validated. In particular, other patients' exams must be collected and corresponding simulations quantitatively evaluated.

In term of decompressed volume, results showed that the large size osteotomy seems to be more efficient from a clinical point of view since, for the same imposed load, it provided a greater decompressed volume and a higher backward displacement (Table 2). Concerning its position, the difference of approximately 5 % between the frontal and back osteotomies is not enough important, compared to the size of the mesh elements, to conclude in a predominance in the position. Nevertheless, the results in term of backward displacement seem to be favourable to a frontal osteotomy since both frontal osteotomies provide greater backward displacement than for osteotomies at the back. These observations seem to point out that a large frontal osteotomy may be optimal.

Table 1 shows that the decompressed volume seems to be not linearly dependant on the size and the location of the osteotomy. Indeed, for an osteotomy twice as small (between the large one of 2.9cm$^2$ and the medium one of 1.4cm$^2$), a decompressed volume of 28 % smaller is produced. Finally, the FE model evaluates the slight forward displacement of the globe that appears after releasing the load.

## 5. Conclusion

The preliminary results obtained for one patient, in accordance with the clinical literature, were satisfying and may help in the improvement of the surgical planning of the exophthalmia reduction: the geometrical model to estimate the decompressed volume (with a estimation error of 3 %) and the FE model to simulate the osteotomy and to compute the resulting backward displacement (with a prediction error of 2 %) and the orbital tissue stiffness (with a estimation error of 19 %). However, it must be kept in mind that (1) the study is based on only one clinical case and (2) that the reliability of quantitative value obtained can be altered by uncertainties of orbital tissue material properties.

As a perspective, thanks to the powerful FE approach, it will be possible to differentiate the anatomical components (optic nerve, muscles, and fat tissue) to rank their role in the global biomechanical behavior of the orbit. The integration of the muscles in the FE model is the next step of our study. In the future, the robustness of the approaches (geometric model and FE model) will be reinforced thanks to a clinical study involving a large base of patients whose data are available at the hospital.

## References

Adenis J.P., Robert P.Y. , 1994. Decompression orbitaire selon la technique d'Olivari. Journal Français d'Ophtalmologie, Vol. 17:pp. 686-691.

Argoubi M., Shizari-Adl A., 1996. Poroelastic creep response analysis of lumbar motion segment in compression. Journal of Biomechanics; Vol. 29: pp.1331-1339.

Biot M.A., 1941. General theory of three-dimensional consolidation. Journal of Applied Physics; Vol.12: pp 155-164.

Buchberger M., Mayr H., 2000. SEE-Kid: software engineering environment for knowledge-based interactive eye motility diagnostics. Proceedings of the Int. Symposium on Telemedicine, Gothenburg, Sweden.

Fung, Y.C., 1993. Biomechanics: Mechanical Properties of Living Tissues. New York: Springer-Verlag.

Jin H.R., Shin S.O., Choo M.J., Choi Y.S., 2000. Relationship between the extent of fracture and the degree of enophthalmos in isolated blowout fractures of the medial orbital wall. Journal of Oral Maxillofacial Surgery; Vol. 58: pp. 617-620.

Miller J. M., Demer J. L., 1999. Clinical applications of computer models for strabismus. Eds Rosenbaum, Santiago, AP, Clinical Strabismus Management. Pub. W. B. Saunders.

Mow V.C., Kuei S.C., Lai W.M., Armstrong C.G., 1980. Biphasic creep and stress relaxation of articular cartilage in compression: theory and experiments. Journal of Biomechanical Engineering, Vol 102: pp. 73-84.

Olivari N., 2001. Endocrine Ophtalmopathy - Surgical treatment. Kaden verlag.

Power E. D., Stitzel J. D., West R. L., Herring I. P., Duma S. M., 2001. A non linear finite element model of the human eye for large deformation loading. Proceedings of the 25$^{th}$ Annual Meeting of Biomechanics, San Diego, 44-45.

Riemann C.D., Foster J.A., Kosmorsky G.S., 1999. Direct orbital manometry in patients with thyroid-associated orbitopathy; vol. 106: pp. 1296-1302.

Sagar M.A., Bullivant D., Mallinson G.D., Hunter P.J., Hunter I.W., 1994. A virtual environment and model of the eye for surgical simulation. Supercomputing; Vol. 2: No. 7.

Saraux H., Biais B., Rossazza C., 1987. Ophtalmologie, Chap. 22 : Pathologie de l'orbite. In : Ed. Mason. : pp. 341-353.

Simon B.R., Wu J.S.S., Carlton M.W., Evans J.H., Kazarian L.E., 1985. Structural models for human spinal motion segments based on a poroelastic view of the intervertebral disk. Journal of Biomechanical Engineering, Vol. 107: pp.327-335.

Stanley R.J., McCaffrey T.V., Offord K.P., DeSanto L.W., 1989. Superior and transantral orbital decompression procedures. Effects on increased intra-orbital pressure and orbital dynamics. Archive of Otolaryngology Head Neck Surgery; vol. 115: pp. 369-373.

Uchio E., Ohno S., Kudoh J., Aoki K., Kisielewicz L.T., 1999. Simulation model of an eyeball based on finite element analysis on a supercomputer. Journal of Ophtalmology. 83(10):1101-2.

Wilson W.B., Manke W.F., 1991. Orbital decompression in Graves' disease. The predictability of reduction of proptosis. Archive of Ophthalmology; vol. 109: pp. 343-345.

*Prediction of tissue decompression in orbital surgery*

Vincent Luboz, Annaig Pedrono, Dominique Ambard, Frank Boutault, Yohan Payan, Pascal Swider

**Captions for illustrations**

**Table 1** – Comparison of the geometrical and FE results with the clinical measurements, for the backward displacement, the volume of decompressed tissue and the stiffness.

**Table 2** – Backward displacement, volume of decompressed tissue and maximum stiffness given by the FE model for a load of 12 N.

**Figure 1 -** Patients suffering from exophthalmia: (a) bilateral exophthalmia (b) bilateral exophthalmia whose only the left eye has been treated: Segmentation of the orbit; (c) Geometric model

**Figure 2 -** Poroelastic FE model of the orbit: **(a)** Medium osteotomy at the back of the orbit ($h_\%$ = 50); **(b)** Large osteotomy at the front of the orbit ($h_\%$ = 62); Greyscale: displacement field of the decompressed tissue out from the osteotomy (free solid displacement and free fluid movement)

**Figure 3 -** Load applied to the globe and resulting backward displacement. After 20s, the equilibrium is reached.

**Figure 4** - Relationships between decompressed tissue volume and backward displacement of the eyeball, derived from the geometric and the FE models.

**Figure 5** - Stiffness of the orbital content (ratio axial load - backward displacement) for 2 sizes and 2 locations of osteotomies ($h_{\%}$=62 % for the front and 50% for the back of the orbit). After 10s, the stiffness decreases to 0 linearly.

*Prediction of tissue decompression in orbital surgery*
Vincent Luboz, Annaig Pedrono, Dominique Ambard, Frank Boutault, Yohan Payan, Pascal Swider

**Table 1 -** Comparison of the geometrical and FE results with the clinical measurements, for the backward displacement, the volume of decompressed tissue and the stiffness.

|  | Clinical results | Geometrical model | Finite Element Model (Large frontal osteotomy) |
|---|---|---|---|
| *Backward displacement*(mm) | 4 | 4 (+0%) | 4.07 (+2%) |
| *Decompressed volume* (cm$^3$) | 3.5 | 3.69 (+3%) | 2.30 (-36%) |
| *Stiffness* (N/mm) | 2.4 | none | 2.85 (+19%) (at t = 7 s) |

*Prediction of tissue decompression in orbital surgery*
Vincent Luboz, Annaig Pedrono, Dominique Ambard, Frank Boutault, Yohan Payan, Pascal Swider

**Table 2** – Backward displacement, volume of decompressed tissue and maximum stiffness given by the FE model for a load of 12 N.

|  | Finite Element Model | | | |
|---|---|---|---|---|
|  | Large osteotomy: 2.9 cm$^2$ | | Medium osteotomy: 1.4cm$^2$ | |
|  | $h_\% = 62$ | $h_\% = 50$ | $h_\% = 62$ | $h_\% = 50$ |
| *Backward displacement* (mm) | 4.07 | 3.39 (-17%) | 2.94 (-28%) | 2.71 (-33%) |
| *Decompressed volume* (cm$^3$) | 2.30 | 2.22 (-3%) | 1.66 (-28%) | 1.76 (-23%) |
| *Maximum stiffness* (N/mm) | 4.27 | 3.24 (-24%) | 6.30 (+48%) | 4.69 (+10%) |

*Prediction of tissue decompression in orbital surgery*
Vincent Luboz, Annaig Pedrono, Dominique Ambard, Frank Boutault, Yohan Payan, Pascal Swider

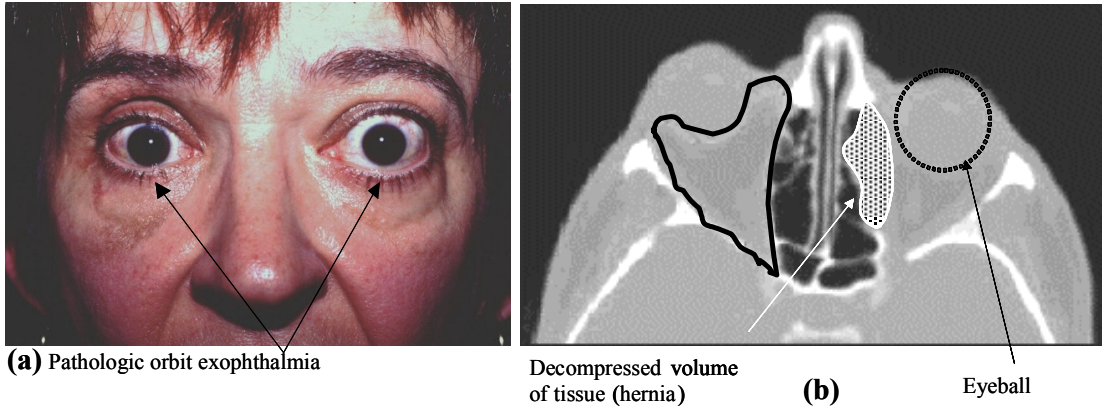

(a) Pathologic orbit exophthalmia

Decompressed volume of tissue (hernia)   (b)   Eyeball

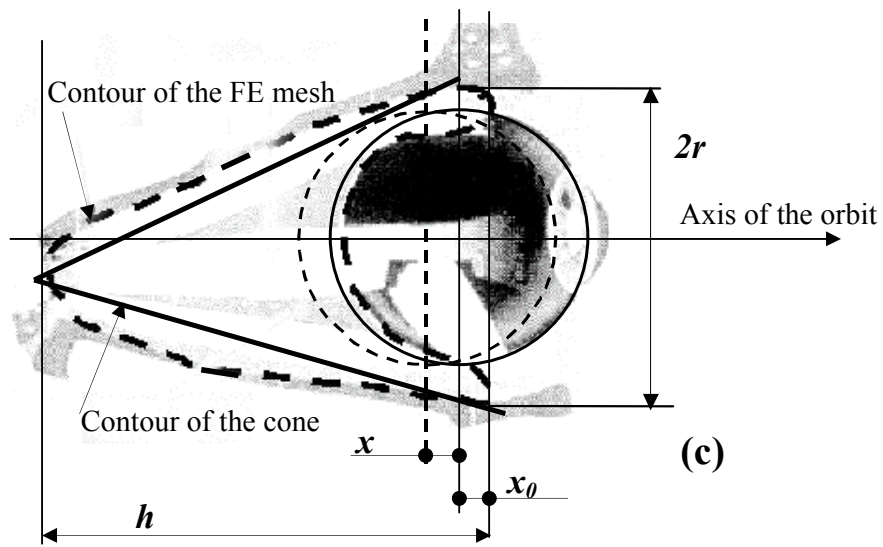

**Figure 1**

*Prediction of tissue decompression in orbital surgery*
Vincent Luboz, Annaig Pedrono, Dominique Ambard, Frank Boutault, Yohan Payan, Pascal Swider

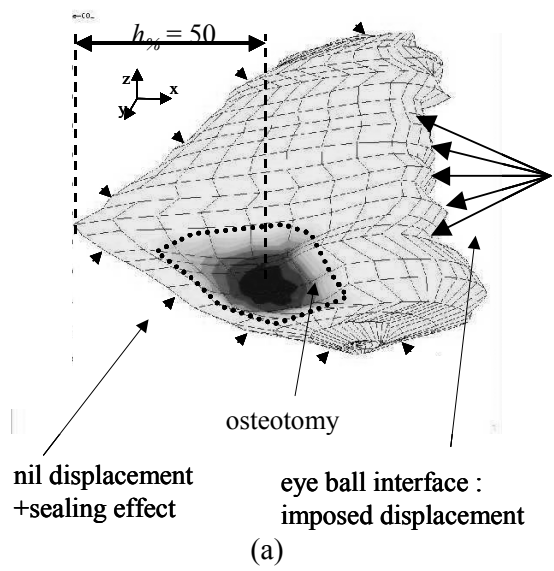

(a)

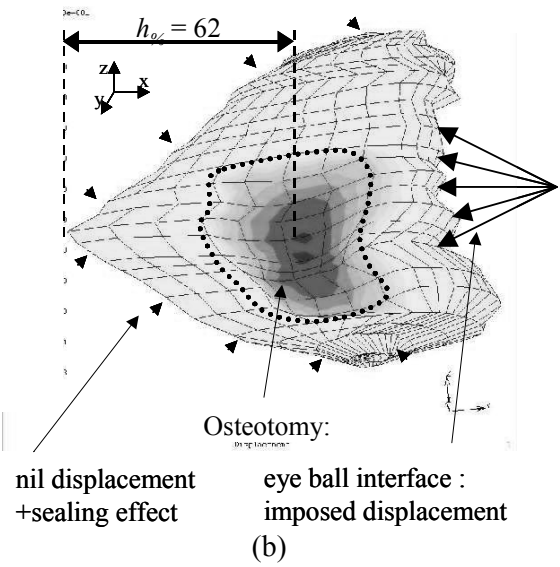

(b)

**Figure 2**

*Prediction of tissue decompression in orbital surgery*
Vincent Luboz, Annaig Pedrono, Dominique Ambard, Frank Boutault, Yohan Payan, Pascal Swider

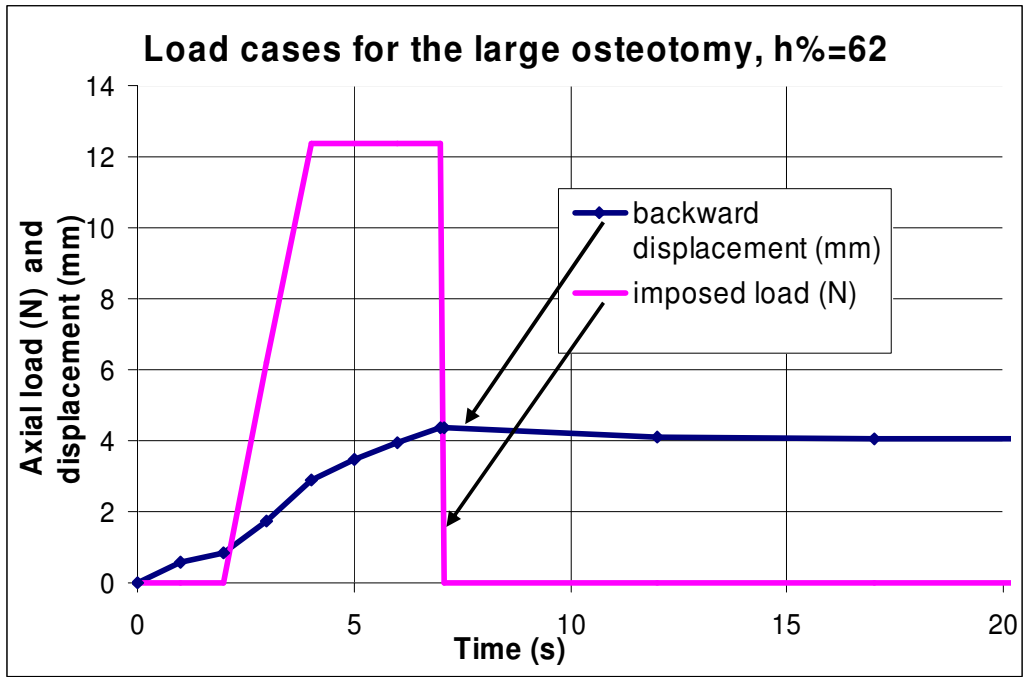

**Figure 3**

*Prediction of tissue decompression in orbital surgery*
Vincent Luboz, Annaig Pedrono, Dominique Ambard, Frank Boutault, Yohan Payan, Pascal Swider

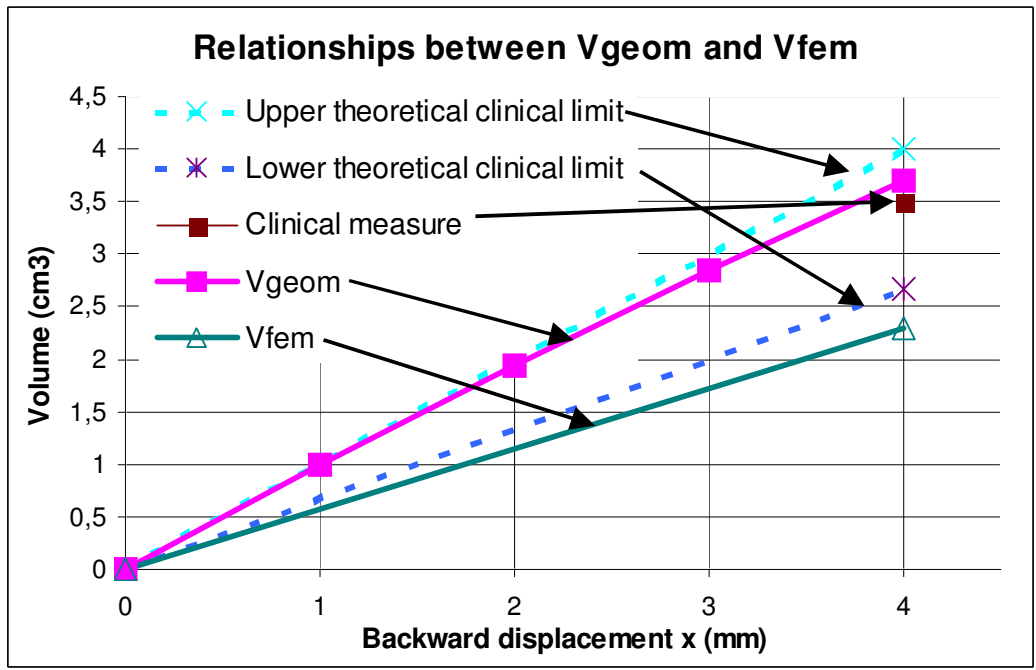

**Figure 4**

*Prediction of tissue decompression in orbital surgery*
Vincent Luboz, Annaig Pedrono, Dominique Ambard, Frank Boutault, Yohan Payan, Pascal Swider

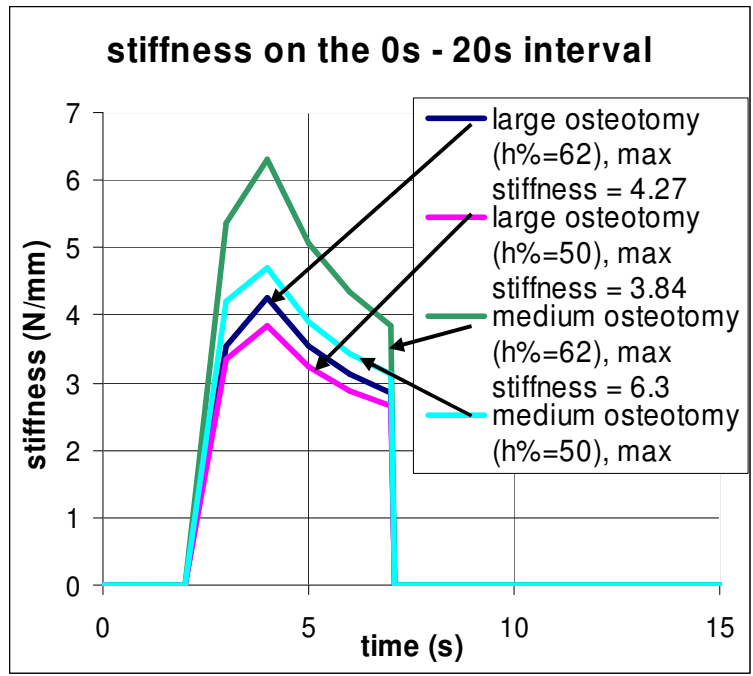

**Figure 5**